\documentstyle[amsmath,amssymb,12pt]{article}
\begin{document}
\setlength{\parskip}{0.45cm}
\setlength{\baselineskip}{0.75cm}
%

%

%
\begin{titlepage}
\setlength{\parskip}{0.25cm}
\setlength{\baselineskip}{0.25cm}
\begin{flushright}
DO-TH 02/02\\
hep--ph/0204090\\
\vspace{0.2cm}
\end{flushright}
\vspace{1.0cm}
\begin{center}
\LARGE
{\bf Precise Ratios for Neutrino-Nucleon}\\ 
\LARGE
{\bf and Neutrino-Nucleus Interactions}
\vspace{1.5cm}

\large
Emmanuel A.\ Paschos\\ \vspace{1.0cm}  
\normalsize
{\it Universit\"{a}t Dortmund, Institut f\"{u}r Physik,}\\
{\it D-44221 Dortmund, Germany} \\
\vspace{0.5cm}

\vspace{1.5cm}
\end{center}

\begin{abstract}
It is shown that several ratios of neutral to charged current cross 
sections are determined accurately, because the largest contribution 
is determined by isospin symmetry and the smaller terms are estimated 
from data or theoretical calculations. This way the theoretical 
uncertainty is very small. 
It is further discussed that the ratios can be 
measured at various distances 
from the origin of the neutrino beams, because an 
increase with distance will be indication for 
neutrino oscillations and will allow a precise determination
of the oscillation parameters.  Finally, coherent
scattering is discussed as a useful reaction for 
oscillations and as a means of searching for new 
light particles.
\end{abstract}
\end{titlepage}


\section{Introduction}

A new generation of neutrino experiments is under construction
and they will be running soon.  The analysis and interpretation
of their results requires accurate calculations of the 
neutrino--nucleon and neutrino--nucleus cross sections.
Several ratios of cross sections are also accessible in 
these experiments: among them are ratios of neutral to charged 
current 
reactions.  It is shown in this article that several ratios
can be calculated precisely, with the largest contribution
obtained from symmetry considerations and smaller terms 
are determined from data and/or theoretical estimates of
reactions.  Having these estimates at our disposal, we
can use them to interpret measurements at various distances
from the origin of the neutrino beams. Changes of
the ratios with the distance from the origin will point to
a change of the neutrino
or antineutrino fluxes and consequently oscillations to
other species of neutrinos.

The article follows the standard model of the electroweak
theory with the charged current given by
\begin{equation}
J_{\mu}^+(x)=(V_{\mu}^1+iV_{\mu}^2)-(A_{\mu}^1+iA_{\mu}^2)
\end{equation}
and neutral current given by
\begin{equation}
J_{\mu}^{NC}=(1-2\sin^2\theta_W)V_{\mu}^3-A_{\mu}^3-
   2\sin^2\theta_W\, V_{\mu}^0\, .
\end{equation}
$V_{\mu}^3$ and the isoscalar part of the current have the quark content
\begin{equation}
V_{\mu}^3 = \frac{1}{2} (\bar{u}\gamma_{\mu}u - \bar{d}\gamma_{\mu}d) \; 
, \,  
V_{\mu}^0=\frac{1}{6}\,(\bar{u}\gamma_{\mu}u
           +\bar{d}\gamma_{\mu}d-2\bar{s}\gamma_{\mu}s)\, , 
\end{equation}
whose contribution to various reactions is determined by
comparing electron induced reactions on protons and 
neutrons.

Ratios of the cross sections were useful 
\cite{ref1}--\cite{ref4}  in establishing lower bounds
for neutral current processes and determined the Weinberg
angle \cite{ref5,ref6}.  In the meanwhile many matrix 
elements of the currents are known.  The experience 
accumulated so far will be used to express ratios of
reactions as equalities.

\section{Cross Sections}
\noindent{\bf{Pion--inclusive Production}}

\noindent Most of the new experiments will be using medium and
heavy nuclei where the interactions of the final pion
are known to be 
important.  We shall consider the scattering of neutrinos
on isospin--zero nuclei and we shall not make the assumption of
single nucleon interactions.  Instead we consider the 
scattering of the current from the complete nucleus 
leading to a final pion and a recoiling system of hadrons
with isospin $I=0,1,2$.  For the isospin analysis we 
adopt the standard notation \cite{ref3} with $U_-^{(I)},
U_0^{(I)}$ being the contributions of the isovector 
current to the cross sections $\sigma_-$ and $\sigma_0$
for charged and neutral current, respectively, and $I$
denotes the isospin of the recoiling hadronic state, i.e.\ the 
states $x_1, \ldots, x_4$ below. 
Similarly, $S_0$ is the isoscalar contribution
to $\sigma_0$ by itself, which occurs only for the $I=1$
final state.  The quantum mechanical structure of the 
cross section gives the relation
\begin{equation}
\frac{\sigma_0^0}{\sigma_-^{ch}-\sigma_-^0} =
\frac{U_0^{(0)}+\frac{2}{5}U_0^{(2)}+\frac{1}{3}S_0}
      {U_-^{(0)}+\frac{2}{5} U_-^{(2)}}
\end{equation}
\vspace{-0.8cm}
with
\begin{eqnarray}
\sigma_-^{ch} & = & 
\sigma \left(\nu _\mu + (I=0) \to \mu^- +\pi^+ +x_1 \right)
   + \sigma\left(\nu_\mu + (I=0) \to \mu^- +\pi^- +x_2\right)\nonumber\\
\sigma_-^0 & = & 
\sigma \left(\nu_\mu + (I=0) \to \mu^-+\pi^0 +x_3\right)\nonumber\\
\sigma_0^0 & = & \sigma \left(\nu_\mu + (I=0) \to \nu_\mu+\pi^0+ x_4\right)
\end{eqnarray}
with $x_1, \ldots ,x_4$ undetected hadronic states.

The corresponding cross sections for antineutrinos 
are obtained by charge symmetry and by changing the sign of the 
vector$\otimes$axial--vector term. Thus, adding neutrino and antineutrino 
reactions eliminates the vector$\otimes$axial--vector
interference term. It follows now 
\begin{equation}
R_0 = \frac{\sigma_0^0+\bar{\sigma}_0^0}
      {(\sigma_-^{ch}-\sigma_-^0)+(\bar{\sigma}_-^{ch}
         -\bar{\sigma}_-^{0})} = \frac{1}{2}\,
\left\{ 1-\frac{\sin^2 2\theta_W \left[V_{em}^{(0)}+V_{em}^{(2)}\right]
   -4\sin^4\theta_W\frac{1}{3}|S_0|^2}
  {\frac{1}{2}
   \left[ (\sigma_-^{ch}-\sigma_-^0)+
     (\bar{\sigma}_-^{ch}-\bar{\sigma}_-^0)\right]}
       \right\}
\end{equation}
with $\theta_W$ the Weinberg angle and 
\begin{equation}
\frac{dV_{em}^{(I)}}{dQ^2d\nu} = \frac{G^2}{\pi}\,
  \frac{Q^4}{4\pi\alpha^2}\, 
       \frac{d\sigma_{em}^{(I)}}{dQ^2d\nu}
\end{equation}
being the isovector contribution of the electromagnetic
current to a final state $x_4$ with isospin $I=0,1,2$.
Similarly, $|S_0|^2$ is the corresponding expression for 
the isoscalar contribution of the electromagnetic current
with only the $I=1$ state contributing to $x_4$.  Since
the ratio within the parenthesis is positive, as will be
shown in eq.\ (15), there is the upper bound
\begin{equation}
R_0 \leq \frac{1}{2}\, .
\end{equation}

\noindent
The ratio in eq.\ (6) is an exact equality.  We have now
sufficient information for neutrino, antineutrino and
electron interactions with nucleons and nuclei to arrive
at an accurate estimate of the right--hand side, without
invoking the inequality.  It will be shown below that the
quotient in the bracket of eq.\ (6) is relatively small so
that even general allowances for the uncertainties still
give an accurate value for $R_0$.

We shall use next information on the electroproduction and
neutrino induced reactions in
order to estimate eq.\ (6). We can complete the electromagnetic
cross section by adding and subtracting the appropriate isoscalar term 
\begin{equation}
R_0 = \frac{1}{2}
 \left[ 1-\sin^2 2\theta_W \frac{V_{em}}
  {\frac{1}{2}(\Sigma+\bar{\Sigma})} +
   (4\sin^4 \theta_W+\sin^2 2\theta_W)\,\frac{1}{3}\,
     \frac{|S_0|^2}{\frac{1}{2}(\Sigma+\bar{\Sigma})})
      \right]\, .
\end{equation}
with $\Sigma=\sigma_-^{ch}-\sigma_-^0$ and a corresponding
expression $\bar{\Sigma}$ for antineutrinos. 
The results so far are general. If we had data for the cross sections 
in nuclei we could substitute them in eq.\ (9). This data is not yet 
available and we shall use cross sections on protons and neutrons.

For the mixing angle we adopt the value $\sin^2\theta_W=
0.2227\pm 0.0004$ \cite{ref5,ref6}.  For $V_{em}$ we use
the data of Galster et al. \cite{ref7} and W.\ Bartel
et al. \cite{ref8} on hydrogen to estimate
\begin{equation}
V_{em} (e+p\to e+\Delta^+) = 
    0.16 \times 10^{-38}\,\,{\rm cm}^2\, 
\end{equation}
at 2.0 GeV. The data show \cite{ref7} that $V_{em}(\pi^0)/V_{em}(ch)$ = 1.5 
and there is at most a $20 \%$ incoherent background. 
The neutrino cross sections on protons and neutrons
were measured in the early experiments:

\begin{center}
\begin{tabular}{|c|c|c|}
\hline
Reaction & Cross section & Reference\\
         & in $10^{-38}$ cm$^2$ & \\
\hline
$\nu p\to\mu^-p\pi^+$ & $0.70\pm 0.10$ & [9,10]\\
$\nu n\to\mu^-p\pi^0$ & $0.20\pm 0.05$ & [9,10]\\
$\nu n\to\mu^-n\pi^+$ & $0.20\pm 0.07$ & [9,10]\\
\hline
$\bar{\nu}n\to\mu^+ n\pi^-$ & 0.31 & theory\\
$\bar{\nu}p\to\mu^+ p\pi^-$ & 0.14 & theory\\
$\bar{\nu}p\to\mu^+ n\pi^0$ & 0.12 & theory\\
\hline
\end{tabular}

Table 1: Neutrino and antineutrino cross sections at $E_\nu = 2$ GeV\@. 
\end{center} 
\noindent
The neutrino cross sections were measured on Hydrogen and Deuterium
where the nucleons are practically free.  The experimental
data were collected together in ref.\ \cite{ref11} 
where 
they are also compared with theoretical calculations.
For the antineutrino reactions there is no data at
low energies and we use theoretical values from 
ref.\ \cite{ref12}. In going from protons and neutrons to nuclei, 
it is necessary to include nuclear effects, which brings in a model 
dependence. 

The new experiments will be using medium
and heavy nuclei as targets where nuclear effects are
important.  Among the corrections are Pauli blocking
at the weak interaction vertex and at the rescatterings, 
absorption of the pions
and charge exchange from the subsequent scatterings.
For nuclear corrections we shall use a model with 
multiple scatterings \cite{ref13} which has been compared 
in a few cases with the experimental data \cite{ref14} 
and was found to be consistent. The charge exchange matrix
has been calculated for several nuclei 
\cite{ref15,ref11}.  We shall consider the scattering
on $_8 O^{16}$ where the charge exchange matrix is \cite{ref11} 
\begin{displaymath} 
M(_8 O^{16}) = A
 \left( \begin{array}{ccc}
   0.78 & 0.16 & 0.06\\
   0.16 & 0.68 & 0.16\\
   0.06 & 0.16 & 0.78\end{array}\right).
\end{displaymath}
The factor $A$ includes the Pauli factor and part of
the absorption.  It is the same for all three 
semileptonic reactions and drops out in the ratios. 
Using the experimental values from Table 1 together
with the mixing matrix we obtain  
\begin{eqnarray}
(\sigma_-^{ch}-\sigma_-^0)_{{\rm Nucl.}\,\,{\rm corr.}}
 &=& A(0.27\pm 0.05)\times 10^{-38}\,\,{\rm cm}^2\\
(\sigma_+^{ch}-\sigma_+^0)_{{\rm Nucl.}\,\,{\rm corr.}}
 & = & A\,\,0.12 \times 10^{-38}\,\,{\rm cm}^2\\
{\rm and}\quad\quad\quad\quad\quad\quad\quad\quad
V_{em} & = & A\,\,0.09\times 10^{-38}\,\,{\rm cm}^2
\end{eqnarray}
for $E_{\nu}=2.0$ GeV.  The values for the antineutrino
cross sections are theoretical and for this reason we 
do not quote any errors.  In the electroproduction 
data the errors are very small.  Using the above 
values we obtain the ratio
\begin{equation}
\sin^2 2\theta_W \frac{V_{em}(\pi^0)}{\frac{1}{2}
   [\Sigma +\bar{\Sigma}]} = 0.70 \cdot
    \frac{0.09}{\frac{1}{2}(0.39)} = 0.32\, .
\end{equation}
The corresponding isoscalar term is small
and is included as a 20\% background to the
electroproduction \cite{ref7}. The final ratio is
\begin{equation}
R_0=\frac{1}{2}(1-0.32 + 0.05) = 0.37
\end{equation}
with the second and third terms coming from the 
isovector and isoscalar part of $V_{em}$, respectively.

Experimental errors and/or other uncertainties stem from
the second and third terms in the parenthesis of eq.\ (9), which 
are much smaller than one.  Thus the overall error 
originating from these terms is reduced; for instance a
20\% error from eq.\ (14) translates to the value
$R_0=0.37\pm 0.03$. The numerical calculation illustrates the 
method to be followed when better data becomes available. 

There are several other ratios which one can discuss, 
but we shall postpone this topic for a more extensive
article and concentrate on different kinds of reactions
which bring out new aspects of the calculations.
An alternative is to use only ratios of neutrino cross
sections. In this case the vector--axial interference
term must be calculated theoretically, 
as was done explicitly in various articles \cite{ref11,ref16}.

\noindent{\bf{Total cross sections}}

\noindent We consider again isoscalar targets and sum over final
states.  In this case there is a popular ratio 
\cite{ref1} where the uncertainties cancel out
\begin{equation}
D=\frac{\sigma_0-\bar{\sigma}_0}{\sigma_- -\bar{\sigma}_-}=
  \frac{1}{2}\, (1-2\sin^2\theta_W)= 0.274\pm 0.002\, .
\end{equation}
This ratio is known precisely and the new experiments can 
measure
the cross sections at various distances from the accelerator,
where the neutrino and antineutrino beams are produced.
An increase of $D$ with the distance is an 
indication for oscillation to active neutrinos, which
provides a model independent method for extracting 
oscillation parameters.  

An alternative ratio is obtained
by adding neutrino and antineutrino cross sections 
\cite{ref1}
\begin{equation}
R_1 = \frac{\sigma_0+\bar{\sigma}_0}{\sigma_-+\bar{\sigma}_-}
  = \frac{1}{2} \left[ 1-\, \frac{\sin^2 2\theta_W\, 
     V_{em}^3 -4\sin^4 \theta_W|J_{em}^0|^2}
    {\frac{1}{2}(\sigma_- +\sigma_+)} \right]
\end{equation}
where $V_{em}^3$ is the isovector contribution (in
the sense of eq.\ (7)) to deep inelastic scattering and
$|J_{em}^0|^2$ the contribution from the isoscalar part
of the current.
We can proceed as in eq.\ (9) to replace $V_{em}^3$ with
the complete electromagnetic contribution.  The
ratio
\begin{equation}
\frac {2 V_{em}}{\sigma_-+\sigma_+} = 0.58 \pm 0.08 
\end{equation}
with the neutrino and antineutrino cross sections taken from the 
particle data group and $V_{em}$ from electroproduction data. 
The value in eq.\ (18) is close to the prediction of the 
parton model \cite{ref17}. Deviations from this value will come from the 
strange and antistrange quark contributions, which are small. For the errors 
we assign a 14 $\%$ error to the electroproduction cross section 
and a 2 $\%$ to the neutrino cross sections to arrive at 
\begin{equation}
R_1 = 0.33 \pm 0.03 
\end{equation}
For the calculation we completed the electromagnetic term by 
adding the isoscalar contribution, which was also subtracted in the 
last term of eq.\ (17), in direct analogy to the method used in eq.\ (9). 
We note that the error in eq.\ (19) is small. 

\noindent{\bf{Coherent Scattering}}

\noindent  
In neutrino experiments, when the final lepton is almost
parallel to the initial neutrino then the leptonic cross
section is related to the pionic one \cite{ref18}
\begin{equation}
\frac{d\sigma}{dQ^2d\nu d\Gamma}= \frac{G^2}{2\pi^2}\,
   \frac{1}{\nu}\, \frac{E'}{E}\, F_{\pi}^2
    \left( \frac{m_{\pi}^2}{m_{\pi}^2+Q^2}\right)^2\,
      \frac{d\sigma_{\pi}}{d\Gamma}\, .
\end{equation}
We adopted the standard notation for the variables with the
phase space element $d\Gamma$ referring to the final hadronic
states and $F_{\pi}$ is the pion decay coupling constant,
$F_{\pi}\approx 0.9 m_{\pi}$. The equation holds for 
neutral and charged current
reactions provided that we restrict the kinematics to the
region where coherent scattering is dominant.  Coherent
scattering occurs when the following conditions are
met:
\begin{enumerate}
\item[(i)] the magnitude of the momentum transfer is
comparable to the square of the mass of the 
particle exchanged, so that a forward peak is produced, 
\item[(ii)] the momentum transfer is small, so that 
the exchanged particle propagates over a region 
comparable to its wave length. If in addition the momentum of the 
incident neutrino is small, its wave function overlaps with several 
scattering centers. 
\end{enumerate}
In coherent scattering the quantum numbers of the region
under investigation add up in the amplitude.
For example, an interaction proportional to the baryon
number counts the baryons contained in the region of
investigation.  Similarly, the electromagnetic interactions
count the charge enclosed in the region of investigation.

A forward peak has been observed in single pion
production by charged currents \cite{ref18} and  the
results were reported to be consistent with the PCAC
hypothesis.  It is possible that other very light particles 
couple directly to the weak currents in a way similar to that 
of the pions. 
Low energy neutrino experiments at the MeV or lower
energies can search for them with the help of coherent scattering. 

\section{Neutral to Charged current ratios in Oscillations}

In the long base line experiments, a beam of
muon--type neutrinos is produced near an accelerator.
As the neutrinos travel a distance $L$, they can 
oscillate to other species of active and/or sterile
neutrinos.  At a distance $L$ the beam is a mixture of
muon--type neutrinos with flux $F_{\mu}(L)$, of another species
of active neutrinos $F_{\nu}(L)$ and, perhaps, of 
sterile neutrinos $F_S(L)$.  The initial condition is
$F_{\mu}(0)=1$ and $F_{\nu}(0)=F_S(0)=0$.  Let us also
assume that the energy of the experiments is low
enough so that, for threshold reasons, the new type of 
active neutrinos do not contribute to the charged current reactions. 
This is the case in the oscillation of muon-- to
tau--neutrinos.  Taking the ratio of neutral to charge
current reactions, one obtains for the number of events
the ratio 
\begin{equation}
\frac{N_{NC}}{N_{CC}} =
\frac{F_{\mu}(L)+F_{\nu}(L)}{F_{\mu}(L)}\cdot
  \frac{\sigma^{NC}}{\sigma^{CC}}\, , 
\end{equation}
which was suggested as a means for analyzing 
oscillation phenomena \cite{ref20}--\cite{ref23}.
The ratio carries two uncertainties:  one coming from
the values of the cross sections and the other from 
the fluxes.
For this reason, the analyses rely on double ratios
\cite{ref24}. The results of this article indicate that for several 
channels the ratio of cross sections is known. This will be useful in the 
analysis of the data and may eliminate the need to use double ratios.

\section{Summary}
We have shown that, in the electroweak theory, ratios of neutral 
to charged current reactions are related to electromagnetic 
reactions and can be estimated in terms of available data. 
The relations (9), (16) and (17) are general and hold at low and high 
energies. They are particularly useful at specific energy regions. 
Eq.\ (9) is more useful in the low energy region $E_\nu < 2.0$ GeV, 
where there are few particles in the final state. Eq.\ (16) has been used 
extensively at various energy regions. Eq.\ (17) is more suitable 
at higher energies where the incoherent scattering from the quarks gives 
the accurate prediction of eq.\ (19). 

The relations were derived for isoscalar targets and hold for 
$I=0$ nuclei as a whole, which eliminates the assumption of single 
nucleon interactions. 
Since electroproduction and antineutrino data is not 
available in nuclei, we used a model for estimating nuclear effects. 
We found out that the final values are rather insensitive to 
experimental uncertainties, because they 
cancel out in ratios and the dominant term follows from isospin 
symmetry. In all cases, the contribution from the isoscalar part 
of the weak neutral current is small. 
We hope that these results will be useful in analyzing 
oscillation experiments. 

The last topic is coherent scattering, which relates neutrino and antineutrino 
induced reactions to the pion induced reactions. In this 
kinematic region the pion induced reactions determine the 
absolute neutrino cross sections. When the square of the 
momentum transfer and the energy transfer are 
small, the exchange current probes a region 
comparable to its wavelength. 
These reactions make possible the searches for light particles that 
couple to neutrinos and hadrons. An interesting case is the coupling 
of neutrinos to the 
baryon current, which will bring a large enhancement. The searches using  
coherent scattering are analogous to the Primakoff effect, which has 
been very useful in the past.

\newpage
\noindent{\large\bf{Acknowledgement}}
The support of the 
``Bundesministerium f\"ur Bildung, Wissenschaft, Forschung und
Technologie'', Bonn under contract 05HT1PEA9 is gratefully 
acknowledged.


\begin{thebibliography}{54}
\bibitem{ref1} E.A.\ Paschos and L.\ Wolfenstein, 
        {\it Phys.\ Rev.} {\bf D7}, 91 (1973)
\bibitem{ref2} A.\ Pais and S.\ Treiman, {\it Phys.\
        Rev.} {\bf D6}, 2700 (1972)
\bibitem{ref3} C.H.\ Albright et al., {\it Phys.\ 
        Rev.} {\bf D7}, 2220 (1973) 
\bibitem{ref4} L.\ Sehgal, {\it Nucl.\ Phys.} 
        {\bf B65}, 141 (1973)
\bibitem{ref5} G.P.\ Zeller et al., 
        {\it Phys.\ Rev.\ Lett.} {\bf 88},
        091802 (2002)
\bibitem{ref6} C.\ Caso et al., Particle Data Group,
        {\it Eur.\ Phys.\ J.} {\bf C15}, 1 (2000)
\bibitem{ref7} S.\ Galster et al.,  
        {\it Phys.\ Rev.} {\bf D5}, 519 (1972)
\bibitem{ref8} W.\ Bartel et al., 
        {\it Phys.\ Lett.} {\bf B28}, 148 (1973) and
        {\it Phys.\ Lett.} {\bf B35}, 181 (1971)
\bibitem{ref9} G.M.\ Radecky et al., {\it Phys.\ Rev.}
        {\bf D25}, 1161 (1982)
\bibitem{ref10} S.J.\ Barish et al., {\it Phys.\ Rev.}
        {\bf D19}, 2521 (1979)
\bibitem{ref11} E.A.\ Paschos, L.\ Pasquali and J.Y. Yu,
        {\it Nucl.\ Phys.} {\bf B588}, 263 (2000)
\bibitem{ref12} J.Y. Yu, Ph.D. Thesis, University of
        Dortmund (2002) 
\bibitem{ref13} S.L.\ Adler, S.\ Nussinov, and E.A.\ Paschos,
        {\it Phys.\ Rev.} {\bf D9}, 2125 (1974)
\bibitem{ref14} P.\ Musset and J.-P. Vialle, 
        {\it Phys.\ Rep.}
        {\bf 39C}, 1 (1978), p.\ 100--101; 
        R.\ Merenyi et al., {\it Phys.\ Rev.} {\bf D45}, 743 (1992)
\bibitem{ref15} S.L. Adler, {\it Phys. Rev.} {\bf D12},
        2644 (1975)
\bibitem{ref16}
C.H. Llewellyn--Smith, {\it Phys.\ Rep.} {\bf 3}, 261 (1972);
G.L. Fogli and G. Nardulli, {\it Nucl.\ Phys.} {\bf B160}, 116 (1979) and 
{\bf B165}, 162 (1980); 
D. Rein and L.M. Sehgal, {\it Annals Phys.} {\bf 133}, 79 (1981); 
E.A.\ Paschos and J.Y. Yu, 
{\it Phys. Rev.} {\bf D65},
        033003 (2002).
\bibitem{ref17} When the cross sections are symmetrized
        over neutrinos and antineutrinos they are given
        by expressions like
        \begin{displaymath}
        \sigma_{\rm tot}=\frac{G^2ME}{\pi} \frac{4}{3}\int F_2(x)dx
        \end{displaymath}
        with $F_2(x)$ the scaling function.  Relations
        among the structure functions of electroproduction
        and neutrino--induced reactions appear in books;
        see, for instance, D.H.~Perkins, Introduction to
        High--Energy Physics, 4th edition (Cambridge University Press), 
eq.\ (5.48) and fig.\ (5.14).
\bibitem{ref18} S.L.\ Adler, 
        {\it Phys. Rev.} {\bf 135}, B963 (1964)
\bibitem{ref19} P.\ Vilain et al., {\it Phys.\ Lett.}
        {\bf B313}, 267 (1993)
\bibitem{ref20} K.\ Nishikawa, INS--Rep.--924 (1992)
\bibitem{ref21} Y.\ Suzuki, in: Proceedings of the Int.\ Conf.\
        on Neutrino Physics and Astrophysics (Neutrino 96)
\bibitem{ref22} F.\ Vissani and A.Y.\ Smirnov, 
        {\it Phys.\ Lett.} {\bf B432}, 376 (1998)
\bibitem{ref23} K.R.S. Balaji et al., {\it Int.\ J.\
        Mod.\ Phys.} {\bf A16}, 1417 (2001)
\bibitem{ref24} C. Mauger, First Int.\ Workshop on
        Neutrino--Nucleus Interactions in the Few GeV
        Region, Dec.\ 13--16, 2001 (KEK, Japan)
\end{thebibliography}
\end{document}